# Topological temporal boundary states

# in a non-Hermitian spatial crystal

Ming-Wei Li[1,2], Jian-Wei Liu[1], Xulong Wang[2], Wen-Jie Chen[1,*], Guancong Ma[2,*], Jian-Wen Dong[1,*]

[1]Department of Physics & State Key Laboratory of Optoelectronic Materials and Technologies, Sun Yat-sen University, Guangzhou 510275, China.

[2]Department of Physics, Hong Kong Baptist University, Kowloon Tong, Hong Kong, China.

*Corresponding author: chenwenj5@mail.sysu.edu.cn, phgcma@hkbu.edu.hk, dongjwen@mail.sysu.edu.cn

**Abstract**

Periodic modulation of the material index in time opens momentum gaps. Such systems are regarded as the temporal analogue of common spatial crystals, wherein the bandgaps open in the frequency space. Recent studies have also led to the theoretical prediction of topological temporal boundary states (TTBSs) in such momentum gaps. In this work, we report the discovery and experimental realization of a new type of TTBS, appearing in a non-Hermitian spatial crystal with spatially periodic loss and gain, wherein the emergence of Bloch momentum gap is associated with a parity-time broken phase, instead of relying on periodic temporal modulation. By inducing a sudden flip of signs of the loss and gain profile, a mode emerges in the middle of the Bloch momentum gap and peaks at the flipping instant, which is regarded as a temporal boundary. Remarkably, we found that the temporal flip induces a topological transition in time, and the said mode is a TTBS that is a temporal analogue of the Jackiw-Rebbi state. The TTBS is experimentally observed in a 1D active mechanical lattice, and it can generically emerge in a wide range of non-Hermitian systems. By linking non-Hermitian physics with spatiotemporal topological systems, our results not only deepen the understanding of temporal topological phases but also open new grounds for controlling transient waves by topological means.

**Main Text**
**Introduction**

Time-varying media represent an emerging horizon for novel wave manipulations[1-14]. Periodic temporal modulation to the material properties of a homogeneous wave-sustaining material (e.g., the dielectric constant), can produce momentum band gaps[10, 12, 15-19]. This phenomenon is analogous to the opening of energy gaps caused by periodic spatial potentials in common crystals[20, 21]. In a momentum gap, modes grow or decay exponentially over time[16, 17, 19, 22-25]. It has been theoretically shown in the context of a photonic time crystal (PTC) that, by judiciously designing the temporal potential profile, the photonic momentum band structure can exhibit distinct topology, depending on the modulation profiles. A temporal boundary can be created at which the modulation profile suddenly changes, driving the system from a topologically trivial to a nontrivial phase. As a consequence of such a temporal topological transition, a topological state peaked at the temporal boundary can be observed[26-31]. Such a topological temporal boundary state (TTBS) is not only conceptually striking, representing a new arena for topological phases, but also potentially useful for robust manipulation of wave signals in the time dimension. Although PTC working in the microwave regime was experimentally realized very recently[32], relevant TTBS has not been observed, with the main challenge being the implementation of a persistent temporal modulation with a sufficient magnitude (comparable to the material index) and rate (at the same order of wave period)[33].

Here, we discover that TTBS can be induced in a non-Hermitian spatial crystal (NSC) by a single temporal flip of its non-Hermiticity. Such a TTBS is found inside a Bloch momentum gap created by the spatial periodic modulation on the imaginary potential (equivalently gain/loss of the material) in the NSC. The Bloch momentum gap here, which corresponds to a parity-time (PT) broken regime[34-47], is distinct from that in PTC, which is caused by temporal modulation to material index. We found that, intriguingly, a topological phase transition of the Bloch momentum band structure can be triggered by a temporal sign flip in the loss and gain profile. A TTBS, which is a temporal analog of a Jackiw-Rebbi state[48, 49], appears and is "localized" at the transition instant. Unlike PTCs, the Bloch momentum gap in our system originates from the spatially periodic modulation of material properties, without resorting to time-dependent changes of material properties, making it both realizable and easily extended to any classical wave system. We experimentally observed the TTBS in a one-dimensional (1D) non-Hermitian mechanical lattice[50-54] and demonstrated its robustness against perturbation and disorder through combined simulations and experiments. This type of TTBS is a momentum-selective phenomenon, existing only at a specific Bloch k point. Contrasting phenomena are also observed in both simulations and experiments. Our work reveals that NSCs with PT-symmetry can serve as viable and more practical platforms for studying spatiotemporal topological phases and enabling topological manipulation of transient signals.

**Results**

**Topological state at the temporal boundary of a non-Hermitian spatial crystal** Consider the 1D spatial crystals supporting classical waves governed by the wave equation. The system can manifest as a photonic, acoustic, or mechanical crystal. As depicted in Fig. 1a, the crystal is a two-layer medium with conjugate constitutive parameters: $\rho_1 = 1 - i\Gamma$ and $\rho_2 = 1 + i\Gamma$. The thicknesses of both layers are identical. The imaginary parts of $\rho_{1,2}$ represent material loss (negative) or gain (positive) and they constitute the non-Hermiticity in the crystal. Figure 1a presents two cases of NSCs, termed LG and GL crystals, respectively, distinguished by their different unit cell configurations. The band structures of these NSCs are calculated by transfer matrix method (see the details in Supplementary Note 1) in Figs. 1b-g for three different values of $\Gamma$. When $\Gamma = 0$, the lattice reduces to a homogenous passive material with a simple linear dispersion (Figs. 1c, 1f). When $\Gamma \neq 0$, this complex NSCs band structures have PT-broken regimes near $k = \pi$, as seen in Figs. 1b, d, e, g. Two complex bands with conjugate eigen frequencies are found in this regime, which correspond to temporally growing (decaying) bulk modes. The temporal behaviors of these modes are similar to the modes inside a momentum gap in a PTC. The difference is that these modes extract (dissipate) energies from (to) material's gain (loss), rather than from (to) temporal modulation of the refractive index[16, 17, 19, 24](see comparison between NSC and PTC in Supplementary Note 4). As rigorously proved below, the PT-broken regime in NSC is indeed a Bloch momentum gap (see Method).

Comparing Figs. 1b-1g, one finds that the band structure undergoes a momentum-gap closing and reopening process as $\Gamma$ changes from positive to negative. This process, which essentially flips the loss and gain in the lattice, swaps the two band edge states, i.e., the EP1 and EP2 modes shown in the insets of Figs. 1b and 1d. In other words, the loss-gain flip resembles band inversion which is a topological transition in 1D topological models. The flip is indeed a topological transition, as we will rigorously show in the next section. We further inspect the eigenmode profiles in the LG (GL) crystals at $k = \pi$, which is inside the Bloch momentum gap. As seen in the insets of Figs. 1e and 1g, the grow mode in the LG crystal and the decay mode in the GL crystal share an identical spatial profile.

Now let us consider the following process. Initially, the grow mode is excited in the LG crystal. At a particular time $t_0$, an instantaneous gain-loss flip is performed by changing the sign of $\Gamma$, such that the LG crystal becomes the GL crystal. This gain-loss flip is essentially a temporal boundary for the system (Fig. 1h), across which the lattice undergoes a topological transition. Fig.1i shows the spatiotemporal profile of the grow and decay modes across the temporal boundary. It is seen that the grow mode in LG crystal is completely coupled to decay mode at $k = \pi$ in GL crystal, forming a TTBS peaked at $t_0$ (see Supplementary Note 3 for details). It is spatially extended while temporally localized at the temporal boundary $t_0$.

**Topological characteristics of the NSCs** The topological characteristics of the NSCs can be revealed by the parallel transport of its eigenmodes[55]. Such a parallel transport is performed along a Bloch momentum band driven by the variation of $\omega$ as a parameter. In other words, the eigenmode here is a function of $\omega$ instead of Bloch $k$. This indicates that the Hamiltonian needs to be written in a form with Bloch $k$ as eigenvalues. To do so, we start from the static case of one NSC (only LG or GL crystal, no temporal flip) shown in Fig.1a. It's governed by a 1D Helmholtz equation:

$$\left[\frac{d^2}{dz^2} + \left(\frac{\omega}{c}\right)^2 \rho(z)\right]\psi(z) = 0,\qquad(1)$$

where $\rho(z) = \begin{cases} 1-i\Gamma & (nd \leq z < nd + d/2) \\ 1+i\Gamma & (nd + d/2 \leq z < (n+1)d) \end{cases}$ is a system-dependent complex constitutive parameter that contains gain and loss, $n$ is the number of the unitcells and c is the speed of wave in a homogeneous system. The eigenmode near the Brillouin zone boundary ($k = \pi$) can be approximately decomposed to $\psi(z) = A(\delta k)E_{+1} + B(\delta k)E_{-1}$ [56], where $\delta k = k - \pi$ is the wavenumber deviated from the $k = \pi$, $E_{+1} = e^{i(\delta k+\pi)z}$ and $E_{-1} = e^{i(\delta k-\pi)z}$ are the +1 and −1 terms of the Fourier series. $A(\delta k)$ and $B(\delta k)$ are the complex amplitudes of the two plane waves. After some algebra (see Supplementary Note 2), we obtain an effective Hamiltonian with $\delta\omega$ as eigenvalues

$$\hat{H}_{\delta\omega}(\delta k) = \alpha(-im_\Gamma \sigma_y - c_3 \delta k \sigma_z),\qquad(2)$$

$$\text{with } H_{\delta\omega}(\delta k)|\psi_{\delta\omega}(\delta k)\rangle = \delta\omega|\psi_{\delta\omega}(\delta k)\rangle.\qquad(3)$$

where $m_\Gamma = \pi^2\sqrt{(\sqrt{\pi^2 + 4\Gamma^2} - \pi)/2\Gamma^2\pi^2}\,\Gamma$, $\delta\omega = \omega - \alpha c_0$. $\alpha$, $c_0$, $c_3$ are constant coefficients. Observe that Eq. (2) can be rewritten as equations with $\delta k$ as eigenvalues (see Method)

$$\hat{H}_{\delta k}(\delta\omega) = -\frac{1}{c_3}\left[\frac{\delta\omega}{\alpha}\sigma_z + m_\Gamma \sigma_x\right],\qquad(4)$$

$$\text{with } H_{\delta k}(\delta\omega)|\psi_{\delta k}(\delta\omega)\rangle = \delta k|\psi_{\delta k}(\delta\omega)\rangle.\qquad(5)$$

The two eigenmomenta ($\delta k$) and the corresponding eigenvectors $|\psi_{\delta k}(\delta\omega)\rangle$ reads

$$\delta k = \pm\frac{1}{c_3}\sqrt{m_\Gamma^2 + \left(\frac{\delta\omega}{\alpha}\right)^2},\qquad(6)$$

$$|\psi_{\delta k}(\delta\omega)\rangle = \frac{1}{N_2}\left[\begin{array}{c} m_\Gamma \\ -\frac{\delta\omega}{\alpha} \mp \sqrt{m_\Gamma^2 + \left(\frac{\delta\omega}{\alpha}\right)^2} \end{array}\right],\qquad(7)$$

where $N_2$ is a normalized coefficient (see Supplementary Note 6). A Bloch momentum gap with $-|m_\Gamma|/c_3 < \delta k < |m_\Gamma|/c_3$ lies between the two EPs. This type of momentum regime with complex conjugate frequencies can also be found in pseudo-Hermitian systems (a generalized form of PT-symmetric system).

They actually act as a Bloch momentum gap, which can be well interpreted using a simple two-band model (see Method).

We are now ready to extract the topological features of the Bloch momentum bands in the LG and GL crystals. Because of the absence of periodicity in $\omega$, here, we perform an open path parallel transport on a Bloch momentum band. Without loss of generality, we discuss the Bloch momentum band on the left side of the Bloch momentum gap ($\delta k < 0$), as shown in Fig. 2a & 2g. Figs. 2b-2f (2h-2l) plot eigenmodes at five representative frequencies in LG (GL) crystal with solid (dash) lines representing the real(imaginary) part. The middle columns (black arrows in colored circles) depict two-component eigenvectors derived from the effective Hamiltonian. From Eq. (7), the eigenvector at the high (low) frequency limit is always $[0,1]^T$ ($[1,0]^T$), regardless of $m_\Gamma$ (see Supplementary Note 6). Notably, the eigenvectors are already nearly parallel to $[1,0]^T$ ($[0,1]^T$) at $0.3(2\pi c/d)$ ($0.7(2\pi c/d)$) in Fig.2 f & l (Fig.2 b & h). These points are then chosen to be the start and finish points of the parallel transport. The same initial gauge, i.e., $[0,1]^T$, is chosen for both the LG and GL crystals (Fig. 2f & 2l). When parallelly transported along the respective Bloch momentum band, the state rotate in opposite directions and evolve into two different EP states at $0.4975(2\pi c/d)$ (Figs. 2d & 2j), whose eigenvectors point to the upper left or upper right direction (blue circles). They both eventually evolve into $[1,0]^T$ at the finish point, but their phases differ by $\pi$ (red circles). This occurs due to the distinct EP modes within left Bloch momentum bands, attributed to the Bloch momentum band inversion (EP1 mode and EP2 mode swap their positions). It's also seen that the two final eigenmodes pick up opposite signs in both real and imaginary parts (Figs. 2b & 2h). This is evident that the LG and GL crystals, which are separated by the Bloch momentum gap-closed case at $\Gamma = 0$, have topologically distinct Bloch momentum bands (see the details in Supplementary Note 6).

In fact, this gap-closing point carries a nontrivial topological charge in 2D synthetic space. To show this, we connect the parallel-transport paths in the LG and GL crystals to form a closed parametric loop in the $\omega - \Gamma$ plane, as shown in Fig. 2m. A degenerate point (DP) with parameters $(\omega d / 2\pi c, \Gamma) = (0.5, 0)$ lies inside the loop. In this $\omega - \Gamma$ plane, the complete picture of the parallelly transported eigenvectors is also plotted in Fig. 2n. With precise calculation, it is revealed that a closed loop integral around the DP (along the red arrow) yields a Berry phase of $\pi$.[20, 57, 58] This is the topological invariant of our system.

**Temporal analogue of Jackiw-Rebbi state at mid-gap momentum** It is quite counterintuitive that a Bloch momentum gap can emerge in an NSC without temporal modulation. Even more interesting is that a topological transition of such an NSC would induce a protected TTBS. Below we will show that the TTBS is, in fact, the temporal analogue of a Jackiw-Rebbi state[59]. To illustrate this, we consider the following dynamic equation based on the effective Hamiltonian in Eq (2): $i\frac{\partial}{\partial t}|\psi(z,t)\rangle = \alpha\left(-im_\Gamma(t)\sigma_y - c_3\delta k\sigma_z\right)|\psi(z,t)\rangle$, where $m_\Gamma(t) = \pi^2 \sqrt{\frac{\sqrt{\pi^6 + 4\Gamma^2\pi^4} - \pi^3}{2\Gamma^2\pi^4}}\Gamma(t)$. For the case in Fig. 1h, $\Gamma(t) = 0.2 step(t-t_0) = \begin{cases} 0.2(t \leq t_0) \\ -0.2(t > t_0) \end{cases}$ is a step-like function indicating a sudden change in the imaginary mass. Thereafter, we find a stable solution at the mid-gap Bloch momentum ($\delta k = 0$) with the wave function given by $|\psi(z,t)\rangle = \begin{cases} \left(A(0)e^{i\pi z} + B(0)e^{-i\pi z}\right)e^{-i\alpha c_0 t}e^{\alpha m_\Gamma(t)*t} & (t \leq t_0) \\ \left(A(0)e^{i\pi z} + B(0)e^{-i\pi z}\right)e^{-i\alpha c_0 t}e^{\alpha m_\Gamma(t)*(2t_0 - t)} & (t > t_0) \end{cases}$. This TTBS consists of a grow bulk mode in the LG crystal ($m_\Gamma(t) > 0$) and a decay bulk mode in the GL crystal ($m_\Gamma(t) < 0$). Furthermore, the TTBS ubiquitously exists universally for arbitrary temporal dependence of $\Gamma(t)$. For a general form of $\Gamma(t)$, the wave function of the TTBS has a Jackiw-Rebbi form[49]:

$$|\psi(z,t)\rangle \propto \left(A(0)e^{i\pi z} + B(0)e^{-i\pi z}\right)e^{-i\alpha c_0 t}e^{\alpha\int^t dt' m_\Gamma(t')} \qquad (8)$$

It implies that the existence of the mid-gap TTBS only requires the sign change in the temporal domains, i.e., $\Gamma(t\rightarrow-\infty) > 0$ and $\Gamma(t\rightarrow\infty) < 0$. We numerically test the robustness of TTBS by simulating different profiles of $\Gamma(t)$. Firstly, we consider a smooth temporal boundary in Fig. 3a, where $\Gamma(t)$ is a sigmoid function. Clearly, this does not affect the existence of the TTBS. The only effect of the smooth boundary is that the temporal peak has a relatively smooth profile. In the second case, we introduce disordered oscillation to $\Gamma(t)$ during the transition between the two crystals (in the time period from $t_0$ to $t_1$). As predicted by Eq. (8), the envelope function (dashed lines in Fig. 3b) first exponentially grows in LG crystal, then oscillates with the disordered variation of $\Gamma(t)$, and finally exponentially decays in GL crystal. Despite these oscillations, the TTBS retains its fundamental structure, highlighting its resilience against temporal disorder. It's also noted that our schemes can in principle apply to any PT-symmetric or pseudo-Hermitian system, such as a ternary/quaternary photonic crystal (see Supplementary Note 9).

**Observation of TTBS in a 1D mechanical lattice** Since our theory is based on a generic wave equation, the TTBS is not restricted to a specific physical system and can emerge in a wide range of PT-symmetric classical wave systems. Here, we use a mechanical lattice consisting of spring-coupled active rotational oscillators[50, 53] to perform a proof-of-principle experiment of the TTBS. The description of the experimental setup is included in the Methods and Supplementary Note 10. Because the TTBS is a spatial bulk state at $k = \pi$, we use two-site unit cells to form a closed 1D chain under Born von-Karman boundary condition. In this configuration, the wave function $|\psi(x)\rangle = [\theta_1, \theta_2, ..., \theta_L]^T$ of this 1D mechanical lattice satisfies $\psi(x_j) = \psi(x_j + L)$, where $\theta_j$ is the measured angular displacement of the $j$ th oscillator site and $x_j$ is its position. Here, we set the total number of sites to be $L = 16$, as shown in Fig.3a. This setup contrasts with traditional open-boundary lattices, as the periodic condition ensures momentum quantization, with Bloch $k = \frac{n\pi}{4}$. Active rotational oscillators can be configured with gain and loss by a self-feedback torque $\tau_j(t) = \beta_j(t)\dot{\theta}_j(t)$, where $\dot{\theta}_j(t)$ is the instantaneous angular velocity of the $j$ th oscillator. $\beta_j(t)$ here determines the onsite loss/gain and is a step function that flips sign at $t_0 = 0$, which corresponds to the non-Hermiticity $\Gamma(t)$ in earlier discussions. The effect of $\tau_j(t)$ can be gain or loss, depending of the sign of $\beta_j(t)$. The amplitude of $\beta_j(t)$ determines the magnitude of gain and loss, which can be tuned at will by the controlling electronics.

To observe the TTBS, the experiment begins by preparing the spatial mode at $k = \pi$ in a passive Hermitian lattice (without loss or gain, i.e. $\beta_j(t) = 0$). This mode is excited by driving the even-numbered oscillators with eight out-of-phase sinusoidal signals at $f = 8.1Hz$ (black arrows in Hermitian crystal in Fig.4b). This excitation profile matches the bipartite eigenmode at $k = \pi$ (see details in Supplementary Note 10). The lattice soon stabilizes into a steady state (white background region in Fig. 4c), which is spatially a bulk mode. Then, the active torques are turned on ($\beta_j(t) = 0 \rightarrow \beta_j(t) > 0$) such that the system transitions to the LG crystal configuration (LG crystal in Fig. 4b), and the oscillations of $\theta_j$ grow exponentially in magnitude, signifying the excitation of the grow mode (pink-shade region in Fig. 4c). After 1.3 s, $\beta_j(t)$ reaches the jump and the system encounters the temporal boundary and flips to the GL crystal configuration ($\beta_j(t) > 0 \rightarrow \beta_j(t) < 0$, GL crystal in Fig. 4b). The oscillations sharply turn from exponentially growth to exponentially decay (green-shade region in Fig. 4c). These observations confirm the existence of TTBS, and they agree well with our theoretical prediction.

**Discussion**

We tested the robustness of TTBS by inserting two time periods (from -0.5s to 0.5s) as "temporal disorder" between LG and GL crystals, as described by $\Gamma(t)$ function in the upper panel of Fig. 4d. The measured displacement fields for all 16 oscillators are plotted in the lower panel of Fig. 4d. It is seen that the existence of the TTBS is robust against the temporal disorder at the time boundary, but its profile is inevitably distorted.

At the first glance, one might tempt to think the TTBS as an apparent phenomenon: wave grows in a gain medium and then decays when the gain is replaced by loss. To show that this is not the case, we conducted a comparison by exciting a spatial mode at $k = 3\pi/4$, which resides within the Bloch momentum gap but deviates from the mid-gap momentum (see Supplementary Note 10). This mode was excited by injecting 16 sinusoidal signals into different sites, and the results are shown in Fig.4e. Unlike the mid-gap mode, the $k = 3\pi/4$ mode exhibited divergent behavior. The displacement field first grows in the LG crystal and then continued to grow despite the configuration is flipped to GL crystal, i.e., the gain in the sites is replaced by loss. The wave in the lattice oscillates and its amplitude eventually diverges. And the temporal peak, where signifies the existence of the TTBS, at $t_0$ is absent. Phenomenologically, the reason for this divergence is that, when deviated from $k = \pi$, the grow mode in LG crystal and the decay mode in GL crystal are mismatched in their spatial profiles. Consequently, the conversion from grow mode to decay mode is incomplete, and some modal components remain growing in the GL crystal and they eventually diverge. These results agree well with our numerical simulation (see the details in Supplementary Note 8).

In summary, we propose a scheme to realize a TTBS in an NSC by flipping the gain and loss at a temporal boundary. The TTBS, which we found to be a temporal Jackiw-Rebbi state, is experimentally observed in an active mechanical lattice. The theoretical model is fundamentally based on a generic wave equation with spatially periodic PT-symmetric distribution of material index as the only requirement, which means our findings are implementable across most classical wave systems, such as electromagnetism, optics, acoustics, and elastic waves. (see Supplementary Note 7 and 8) The TTBS reported here relies only on the interplay of spatial crystal and the temporal flip of non-Hermicity, which is unlike the TTBS appearing in PTC that demands persistent temporally periodic modulation to material indices, making it relatively straightforward to implement. Our work not only extends the concept of topological phases for spatiotemporal scenarios but also offers new routes for the spatial and transit control of waves. Momentum selective feature of TTBS may be used as a momentum filter, which is complementary to traditional frequency filters.

**Materials and Methods**

**Bloch momentum gap induced by imaginary mass in PT system** In this section, we will prove that viewing a PT-broken regime as a Bloch momentum gap is reasonable. For ease of understanding, we first look at a simple two-band model, written as $\hat{H}_\omega(k) = k\sigma_z$, which satisfies $\hat{H}_\omega(k)\psi = \omega\psi$. It has a gapless dispersion with a linear band crossing at zone center, namely a diabolic point. Upon adding a real mass $m_R\sigma_y$ to the Hamiltonian $\hat{H}_\omega(k) = m_R\sigma_y + k\sigma_z$, a frequency bandgap opens, which is commonly found in Hermitian systems. Inside the band gap exist two branches of eigenmodes with real ω and complex $k$, which evanescently decay or grow in space. Notably, systems with real masses of opposite signs produce topologically distinct gaps, expecting a protected interface mode localized at their domain wall[59]. Good examples are Su–Schrieffer–Heeger lattices[60] and their high-dimensional counterparts[61]. By contrast, one can introduce an imaginary mass $im_I\sigma_y$ to the diabolical Hamiltonian, leading to a non-Hermitian Hamiltonian:

$$\hat{H}_\omega(k) = im_I\sigma_y + k\sigma_z, \tag{9}$$

which has a similar form to Eq. (2). Inside the momentum regime $k \in (-|m_I|, |m_I|)$, which is usually called PT-broken regime, resides two branches of eigenmodes with complex frequencies, as is common in non-Hermitian systems with PT symmetry. The matrix form of eigen equation reads

$$\begin{pmatrix} k & m_I \\ -m_I & -k \end{pmatrix} \begin{pmatrix} p_1 \\ p_2 \end{pmatrix} = \omega \begin{pmatrix} p_1 \\ p_2 \end{pmatrix}, \tag{10}$$

where $\psi = [p_1, p_2]^T$, the equations can be expanded as

$$\begin{aligned} kp_1 + m_I p_2 &= \omega p_1 \\ -m_I p_1 - k p_2 &= \omega p_2 \end{aligned} \tag{11}$$

They are equivalent to:

$$\omega p_1 - m_I p_2 = kp_1$$
$$-m_I p_1 - \omega p_2 = kp_2 \quad (12)$$

which is an eigen equation for momentum k $\hat{H}'_k(\omega)\psi = k\psi$ with $\hat{H}'_k(\omega) = -m_I \sigma_x + \omega \sigma_z$. After a unitary transformation $U\hat{H}'_k(\omega)U^{-1}$, we have

$$\hat{H}_k(\omega) = m_I \sigma_y + \omega \sigma_z \quad (13)$$

It treats Bloch momentum k as eigenvalue and takes a similar form to $\hat{H}_\omega(k) = m_R \sigma_y + k\sigma_z$, except that frequency $\omega$ and Bloch momentum k swap roles. In this regard, the PT-broken regime can be considered as a gap opening in the Bloch momentum axis. Moreover, a topologically protected mode is expected to be localized at the domain wall between two crystals with opposite imaginary mass, which is exactly the TTBS we discussed in the main text.

Bloch momentum gap in our system is equivalent to k gap in PTCs under two-band approximation, although they are opened by different mechanisms. We analytically derived the Hamiltonian of the PTC near its k gap using frequency decomposition (see the details in Supplementary Note 5),

$$\hat{H}_k(\omega) = m_I \sigma_y + c_4 \delta\Omega \sigma_z, \quad (14)$$

where $\delta\Omega$ is the Floquet frequency deviated from mid gap and $m_I$ is the Dirac mass measuring the temporal modulation strength in PTC. It suggests that the wave behaviors in k gap of a PTC should be analogous to that in the Bloch momentum gap of a NSC.

To compare their wave behaviors, we simulated the propagation of a Gaussian pulse under the influence of an NSC (see the details in Supplementary Note 5). In this simulation, the temporal reflection, refraction and amplification are observed, when Bloch momentum resides inside or outside the gap, which is similar to the phenomena reported in PTC[26].

**The experimental setup** We utilize a lattice of coupled active rotational harmonic oscillators for the experiments[53]. The building block is constructed using a brushless DC motor (specifically, the LDPOWER 2804 model), to which a cross-shape rotating arm is affixed. The arms measure 129.64 mm in length and 93.58 mm in width. Two large holes are drilled on the ends of the arms to accommodate the insertion of weights for the adjusting of the moment of inertia. Two identical anchoring springs are affixed to the arms to provide restoring torques, each with a spring constant of 32.12 N/m. The resonance frequency of an isolated oscillator is determined by the spring constant and the moment of inertia. By either changing the moment of inertia (by the loaded weights) or the restoring torque (the tension in the springs or the force arm), different resonant frequency can be achieved. The response spectra for sites 1 to 16 have been recorded, with a measured resonance frequency of 8.1 Hz (Supplementary Note 10). To form a lattice structure, we link multiple oscillators using extra tensioned springs to join the adjacent units. A diametrically magnetized magnet is firmly attached beneath the shaft of the motor, and a magnetic rotary position sensor (AMS AS5047P) is placed 2mm below the magnet to measure the instantaneous angular displacement $\theta(t)$ of each oscillator. The measured signals are digitized at a sampling rate of 1 kHz by a microcontroller (Espressif ESP32) and sent to a computer. The recorded signals are the local response. They are also utilized as feedback for the generation of gain and loss. To achieve a PT-symmetric crystal, the onsite gain and loss are induced by exerting angular-velocity-dependent torque $\tau(t) = \beta(t)\dot{\theta}(t)$ to the electric motor via an active feedback-control loop[50]. The angular velocity $\dot{\theta}(t)$ is calculated from the angular displacement by a differential algorithm. When the value of $\beta$ is positive (negative), the influence of $\tau(t)$ is correspondingly gain (loss). Specified instantaneous torque based on the angular velocity is programed to the oscillator's motor with a driver chip (STMicroelectronics L6234PD) in the microcontroller. To realize the PT-symmetric condition, we aim to establish a balance between gain and loss. Given that the system exhibits an intrinsic loss of $Im(f)$ =-0.23Hz (see the details of measurements in Supplementary Note 10), a stronger torque should be applied to the gain sites compared to that at the loss sites. In the experiment shown in Fig. 4c & 4e, we chose the value of G (L) to be $Im(f) = 0.66$Hz (-0.20Hz).

**Acknowledgments**
M. L. and G. M. thank Mathias Fink for discussion. This work was supported by National Key R&D Program of China (Grant Nos. 2022YFA1404304, 2019YFA0706302, 2022YFA1404400), Guangdong Basic and Applied Basic Research Foundation (Grant No. 2023B1515040023), and the Hong Kong Research Grants Council (RFS2223-2S01, 12301822).

**References**
1. B. A. AULD, J.H.C., H. ROLAND ZAPP, *Signal processing in a nonperiodically time-varying magnetoelastic medium.* IEEE Antennas and Propagation Magazine, 1968. **56**(3).
2. Felsen, L. and G. Whitman, *Wave propagation in time-varying media.* IEEE Transactions on Antennas and Propagation, 1970. **18**(2): p. 242-253.
3. J.T.Mendonca, P.K.S., *Time Refraction and Time Reflection: Two Basic Concept.* Physica Scripta, 2002. **65**: p. 160-163.
4. Bacot, V., et al., *Time reversal and holography with spacetime transformations.* Nature Physics, 2016. **12**(10): p. 972-977.
5. Camacho, M., B. Edwards, and N. Engheta, *Achieving asymmetry and trapping in diffusion with spatiotemporal metamaterials.* Nature Communications, 2020. **11**(1): p. 3733.
6. Li, H. and A. Alù, *Temporal switching to extend the bandwidth of thin absorbers.* Optica, 2020. **8**(1).
7. Pacheco-Peña, V. and N. Engheta, *Temporal aiming.* Light: Science & Applications, 2020. **9**(1): p. 129.
8. Pacheco-Peña, V. and N. Engheta, *Antireflection temporal coatings.* Optica, 2020. **7**(4).
9. Li, H., et al., *Temporal Parity-Time Symmetry for Extreme Energy Transformations.* Phys Rev Lett, 2021. **127**(15): p. 153903.
10. Galiffi, E., et al., *Photonics of time-varying media.* Advanced Photonics, 2022. **4**(01).
11. Tirole, R., et al., *Saturable Time-Varying Mirror Based on an Epsilon-Near-Zero Material.* Physical Review Applied, 2022. **18**(5): p. 054067.
12. Yin, S., E. Galiffi, and A. Alù, *Floquet metamaterials.* eLight, 2022. **2**(1): p. 8.
13. Hidalgo-Caballero, S., et al., *Damping-Driven Time Reversal for Waves.* Physical Review Letters, 2023. **130**(8): p. 087201.
14. Moussa, H., et al., *Observation of temporal reflection and broadband frequency translation at photonic time interfaces.* Nature Physics, 2023. **19**(6): p. 863-868.
15. Reyes-Ayona, J.R. and P. Halevi, *Observation of genuine wave vector (k or β) gap in a dynamic transmission line and temporal photonic crystals.* Applied Physics Letters, 2015. **107**(7).
16. Lyubarov, M., et al., *Amplified emission and lasing in photonic time crystals.* Science, 2022. **377**(6604): p. 425-428.
17. Sharabi, Y., et al., *Spatiotemporal photonic crystals.* Optica, 2022. **9**(6): p. 585-592.
18. Lustig, E., et al., *Photonic time-crystals - fundamental concepts [Invited].* Optics Express, 2023. **31**(6): p. 9165-9170.
19. Pan, Y., M.-I. Cohen, and M. Segev, *Superluminal k-Gap Solitons in Nonlinear Photonic Time Crystals.* Physical Review Letters, 2023. **130**(23): p. 233801.
20. Bernevig, B.A., *Topological Insulators and Topological Superconductors*. 2013, Princeton: Princeton University Press.
21. Ozawa, T., et al., *Topological photonics.* Reviews of Modern Physics, 2019. **91**(1).
22. Caloz, C. and Z.L. Deck-Léger, *Spacetime Metamaterials—Part II: Theory and Applications.* IEEE Transactions on Antennas and Propagation, 2020. **68**(3): p. 1583-1598.
23. Lee, S., et al., *Parametric oscillation of electromagnetic waves in momentum band gaps of a spatiotemporal crystal.* Photonics Research, 2021. **9**(2): p. 142-150.
24. Sharabi, Y., E. Lustig, and M. Segev, *Disordered Photonic Time Crystals.* Phys Rev Lett, 2021. **126**(16): p. 163902.
25. Sadhukhan, S. and S. Ghosh, *Bandgap engineering and amplification in photonic time crystals.* Journal of Optics, 2024. **26**(4): p. 045601.
26. Lustig, E., Y. Sharabi, and M. Segev, *Topological aspects of photonic time crystals.* Optica, 2018. **5**(11).
27. Ma, J. and Z.G. Wang, *Band structure and topological phase transition of photonic time crystals.* Opt Express, 2019. **27**(9): p. 12914-12922.
28. Dong, R.Y., et al., *Band Structure and Temporal Topological Edge State of Continuous Photonic Time Crystals.* IEEE Transactions on Antennas and Propagation, 2024. **72**(1): p. 674-682.
29. Feis, J., et al., *Spacetime-topological events.* arxiv, 2024.
30. Lin, M., et al., *Temporally-topological defect modes in photonic time crystals.* Optics Express, 2024. **32**(6).
31. Zhu, W. and J.-h. Jiang, *Characterizing generalized Floquet topological states in hybrid space-time dimensions.* arxiv, 2024.
32. Wang, X., et al., *Metasurface-based realization of photonic time crystals.* 2023. **9**(14): p. eadg7541.

33. Boltasseva, A., V.M. Shalaev, and M. Segev, *Photonic time crystals: from fundamental insights to novel applications: opinion.* Optical Materials Express, 2024. **14**(3): p. 592-597.
34. Bender, C.M. and S. Boettcher, *Real Spectra in Non-Hermitian Hamiltonians Having PT Symmetry.* Physical Review Letters, 1998. **80**(24): p. 5243-5246.
35. Mostafazadeh, A., *Pseudo-Hermiticity versus PT symmetry: The necessary condition for the reality of the spectrum of a non-Hermitian Hamiltonian.* Journal of Mathematical Physics, 2002. **43**(1): p. 205-214.
36. Bender, C.M., *Making sense of non-Hermitian Hamiltonians.* Reports on Progress in Physics, 2007. **70**(6): p. 947-1018.
37. El-Ganainy, R., et al., *Theory of coupled optical PT-symmetric structures.* Optics Letters, 2007. **32**(17): p. 2632-2634.
38. Klaiman, S., U. Günther, and N. Moiseyev, *Visualization of Branch Points in $\mathcal{P}\mathcal{T}$-Symmetric Waveguides.* Physical Review Letters, 2008. **101**(8): p. 080402.
39. Makris, K.G., et al., *Beam Dynamics in $\mathcal{P}\mathcal{T}$ Symmetric Optical Lattices.* Physical Review Letters, 2008. **100**(10): p. 103904.
40. Guo, A., et al., *Observation of $\mathcal{P}\mathcal{T}$-Symmetry Breaking in Complex Optical Potentials.* Physical Review Letters, 2009. **103**(9): p. 093902.
41. Rüter, C.E., et al., *Observation of parity–time symmetry in optics.* Nature Physics, 2010. **6**(3): p. 192-195.
42. Regensburger, A., et al., *Parity–time synthetic photonic lattices.* Nature, 2012. **488**(7410): p. 167-171.
43. Feng, L., R. El-Ganainy, and L. Ge, *Non-Hermitian photonics based on parity–time symmetry.* Nature Photonics, 2017. **11**(12): p. 752-762.
44. El-Ganainy, R., et al., *Non-Hermitian physics and PT symmetry.* Nature Physics, 2018. **14**(1): p. 11-19.
45. Zhao, H. and L. Feng, *Parity–time symmetric photonics.* National Science Review, 2018. **5**(2): p. 183-199.
46. Ozdemir, S.K., et al., *Parity-time symmetry and exceptional points in photonics.* Nat Mater, 2019. **18**(8): p. 783-798.
47. Gupta, S.K., et al., *Parity-Time Symmetry in Non-Hermitian Complex Optical Media.* Adv Mater, 2020. **32**(27): p. e1903639.
48. Jackiw, R. and C. Rebbi, *Solitons with fermion number 1/2.* Physical Review D, 1976. **13**(12): p. 3398-3409.
49. Shen, S.Q., *Topological Insulators_ Dirac Equation in Condensed Matter* Springer Series in Solid-State Sciences. 2017: Springer Science & Business Media.
50. Wang, W., et al., *Experimental Realization of Geometry-Dependent Skin Effect in a Reciprocal Two-Dimensional Lattice.* Physical Review Letters, 2023. **131**(20): p. 207201.
51. Cui, X., et al., *Experimental Realization of Stable Exceptional Chains Protected by Non-Hermitian Latent Symmetries Unique to Mechanical Systems.* 2023. **131**: p. 237201.
52. Wang, W., X. Wang, and G. Ma, *Extended State in a Localized Continuum.* Physical Review Letters, 2022. **129**(26): p. 264301.
53. Wang, W., X. Wang, and G. Ma, *Non-Hermitian morphing of topological modes.* Nature, 2022. **608**(7921): p. 50-55.
54. Nash, L.M., et al., *Topological mechanics of gyroscopic metamaterials.* PNAS, 2015. **112**(47): p. 14495-14500.
55. Raffaele, R., *Manifestations of Berry's phase in molecules and condensed matter.* Journal of Physics: Condensed Matter, 2000. **12**(9): p. R107.
56. Ding, K., Z.Q. Zhang, and C.T. Chan, *Coalescence of exceptional points and phase diagrams for one-dimensional PT-symmetric photonic crystals.* Phys. Rev. B., 2015. **92**(23).
57. Raffaele, R., *Manifestations of Berry's phase in molecules and condensed matter.* J. Phys.: Condens. Matter, 2000. **12**(9): p. R107.
58. Vanderbilt, D., *Berry Phases in Electronic Structure Theory: Electric Polarization, Orbital Magnetization and Topological Insulators.* 2018, Cambridge: Cambridge University Press.
59. János K. Asbóth , L.O., András Pályi, *A Short Course on topological insulator.* 2015.
60. Su, W.P., J.R. Schrieffer, and A.J. Heeger, *Solitons in Polyacetylene.* Physical Review Letters, 1979. **42**(25): p. 1698-1701.
61. Dutt, A., et al., *Higher-order topological insulators in synthetic dimensions.* Light: Science & Applications, 2020. **9**(1): p. 131.

**Figures and Tables**

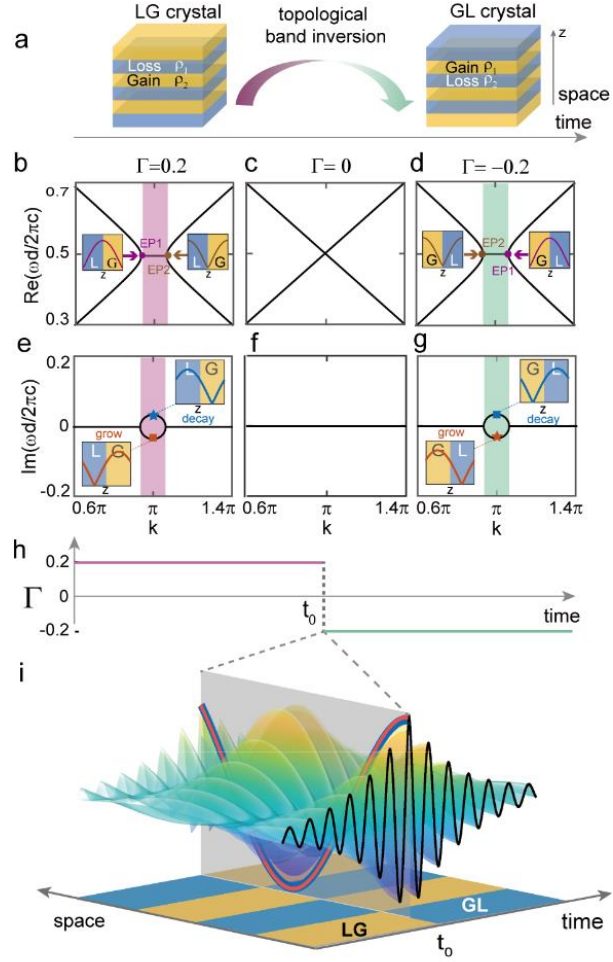

**Figure 1. TTBS between two NSC. a** Topological band inversion induced by non-Hermiticity flipping in a 1D binary crystal. Unit cells of both crystals consists of two layered media with mutually conjugate constitutive parameters ( $\rho_1 = 1-i\Gamma$ and $\rho_2 = 1+i\Gamma$ ). LG crystal: $\Gamma=0.2$ , GL crystal: $\Gamma=-0.2$ . **b-g** Complex band structures for the cases of $\Gamma=0.2, 0, -0.2$ . As non-Hermiticity $\Gamma$ flips sign, the band structure experiences a topological band inversion, along with two EPs swapping their positions. **h** Temporal function of $\Gamma$ indicating a topological phase transition occurred in time $t_0$ . **i** Spatiotemporal profile of the TTBS. It is localized at the temporal boundary $t_0$ , which is comprised of the mid-gap grow mode in LG crystal ( $t < t_0$ ) and the mid-gap decay mode in GL crystal ( $t > t_0$ ). The red and blue line represent the wave function at $t = t_0^-$ in LG crystal and $t = t_0^+$ in GL crystal.

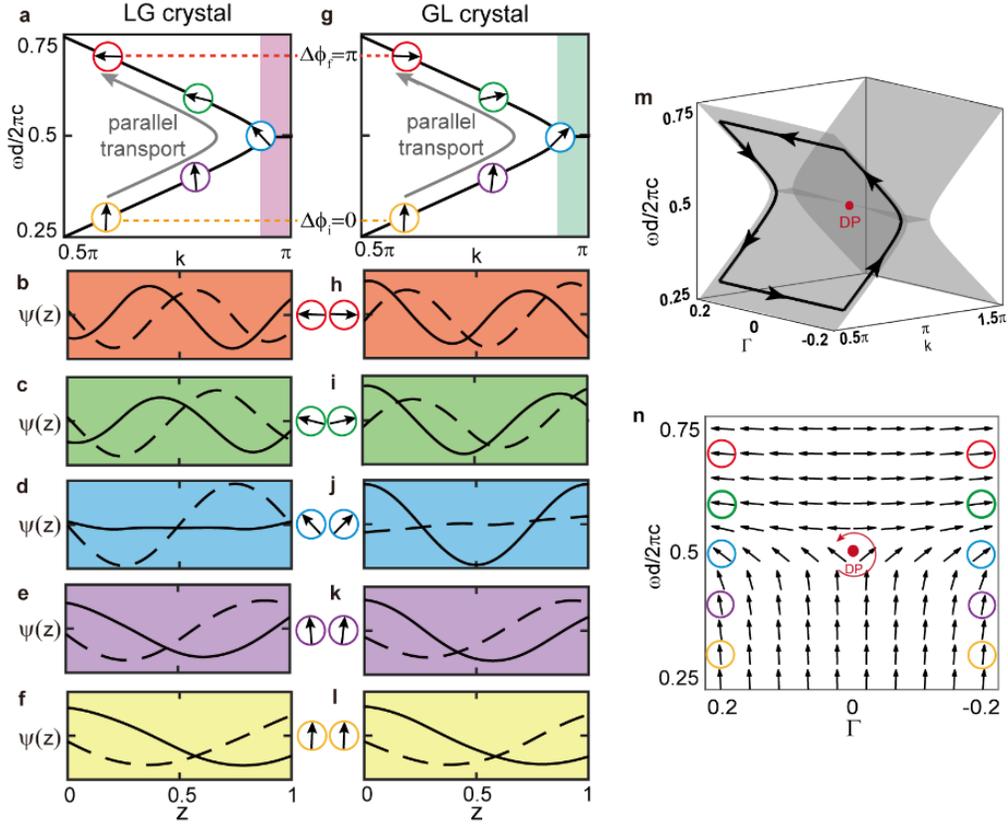

**Figure 2. Parallel transport of the eigenmodes along the momentum bands. a & g** Real band structures for LG and GL crystals on the left side of the gap. **b–f** Eigenmodes at five representative frequencies for LG crystal. Solid lines and dash lines represent their real and imaginary part. **h–l** Eigenmodes for GL crystal. The gray arrows in a and g indicate the directions of parallel transport. The two eigenvectors at frequency of $0.3(2\pi c/d)$ (yellow circles) are chosen as the initial states of parallel transport, for which the same gauge has been chosen (f and l). Using parallel transport gauge, we find that the final states in b and h accumulate a phase difference of $\pi$, as predicted by the eigenvectors of effective Hamiltonians (arrows in b–l). This indicates the distinct topologies of momentum bands in the two crystals. **m.** A closed path defining Berry phase in ω-Γ synthetic space. **n.** The complete picture of the parallel transport eigenvectors in ω-Γ plane, with color circles corresponding to b-l. Integral along the red arrow in a closed loop, enclosing the DP, can derive a Berry phase of $\pi$.

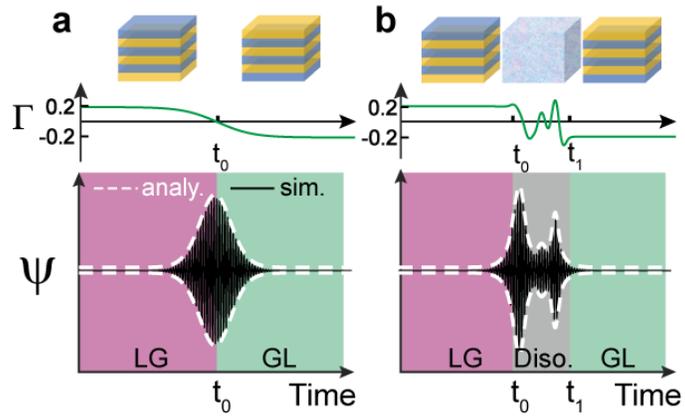

**Figure 3. Robust boundary states between two topologically distinct NSCs.** To demonstrate the ubiquitous existence of TTBS, we simulate two cases of temporal boundaries. **a** Temporal boundary between LG crystal ($\Gamma = 0.2$) and GL crystal ($\Gamma = -0.2$) with a smooth variation in $\Gamma$. **b** Temporal boundary with a disordered variation of $\Gamma$ in between the two crystals.

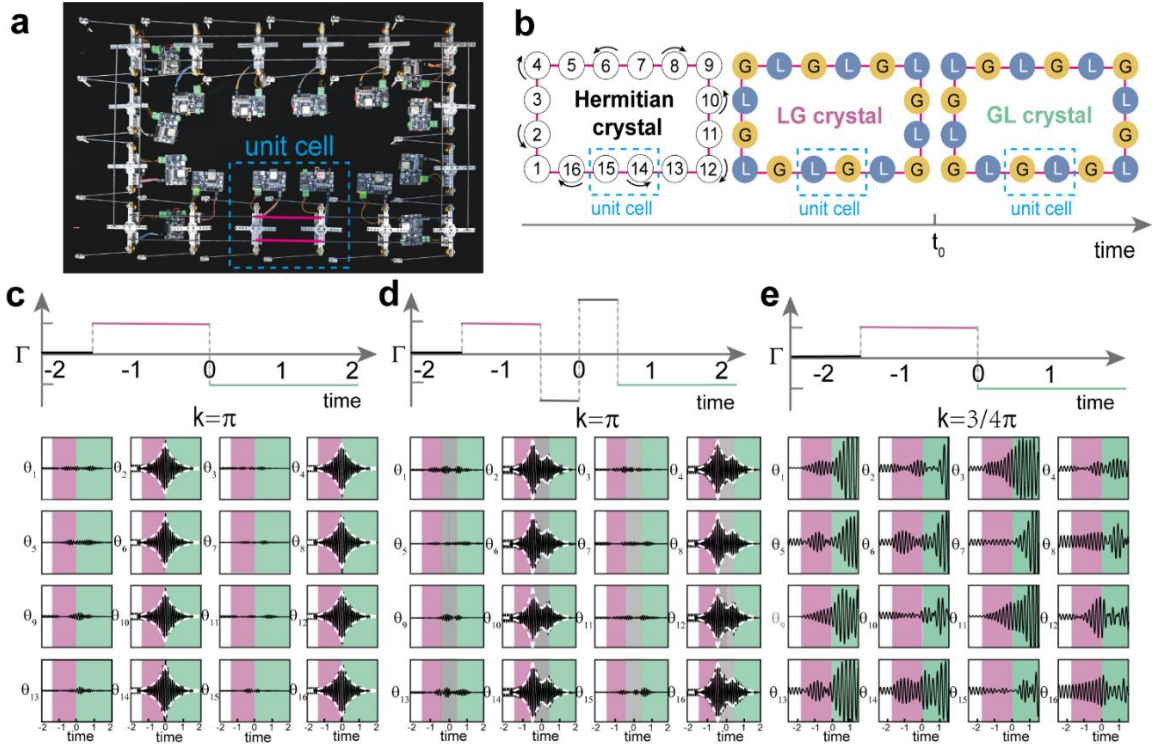

**Figure 4. Experimental observation of TTBS in 1D mechanical lattices**. **a** Experimental setup of the active mechanical lattice. Its unit cell (blue dashed box) consists of two mechanical oscillators (one with gain and the other with loss). Sixteen oscillators are connected by coupling strings from head to tail, satisfying the Born von-Karman boundary condition. **b** Three time periods in exciting TTBS. **c** Measured angular displacements of the TTBS between LG and GL crystals. The variation of non-Hermiticity $\Gamma(t)$ is plotted in upper panel. The time period of LG (GL) crystals shaded in pink (green). **d** Measured angular displacements of the TTBS of a temporal boundary with disorder. In **c** & **d**, the systems are excited by a driving signal with $k = \pi$, consistent with the Bloch $k$ of TTBS. **e** Measured angular displacements when the system is initially excited by a driving signal with $k = 3\pi/4$. The measured displacements eventually diverged in GL crystal and the TTBS peak disappears near the temporal boundary, verifying the momentum selective feature.